\documentclass[
    ,final
  ,numberedheadings 
  ,cmfonts        
  ]
  {aipproc}

\usepackage{amsmath}
\usepackage{subfigure}
\layoutstyle{6x9}

\newcommand{\ce}[1]{Eq.~(\ref{#1})}

\newcommand{\cf}[1]{{Fig.~\ref{#1}}}
\newcommand{\ct}[1]{{Tab.~\ref{#1}}}
\newcommand{\ep}{\varepsilon}

\newcommand{\vect}[1]{\vec{#1}}

\newcommand{\eqs}[1]{\begin{equation} \begin{split} #1\end{split} \end{equation} }

\newcommand{\ks}[1]{#1 \!\!\!\!\! \slash }

\newcommand{\gmu}{\gamma^\mu}
\newcommand{\gnu}{\gamma^\nu}

\newcommand{\ie}{{\it i.e.}}
\newcommand{\eg}{{\it e.g.}}

\renewcommand{\thefootnote}{\fnsymbol{footnote}}

\begin{document}

\title{Ambiguities in the calculation of leptonic decays of excited heavy quarkonium}


\author{J.P.~Lansberg}{
address={Physique th\'eorique fondamentale, D\'epartement de  Physique, Universit\'e de  Li\`ege, \\ all\'ee du 6 Ao\^{u}t 17, b\^{a}t. B5, B-4000 Li\`ege~1, Belgium\\
E-mail: JPH.Lansberg@ulg.ac.be}
}

\begin{abstract}
We point out that the determination of the leptonic decay width of radially-excited quarkonia
is strongly dependent on the position of the node typical of these excitations.
We suggest that this feature could be related with the longstanding  $\rho-\pi$ puzzle.

\end{abstract}

\maketitle


\footnotetext{Presented at HADRONIC PHYSICS (HLPR 2004): Joint Meeting Heidelberg-Li\`ege-Paris-Rostock, Spa, Belgium, 
16-18 December 2004.}
\renewcommand{\thefootnote}{\arabic{footnote}}

\section{Introduction}

To study processes involving heavy quarkonia, 
such as decay and production mechanisms, in field theory,  all the information
needed can be parameterised by vertex functions, which describe
the coupling of the bound state to its constituents and contain the information about the size of 
the bound state, the amplitude of probability for given quark configurations and
the normalisation of the bound-state wave functions. 

In a parallel work~\cite{these,Lansberg:2005pc,article2}, we have have considered production processes of
$J/\psi$, $\psi'$ and $\Upsilon$ at hadron colliders. We relied on a phenomenological approach 
with vertex functions, and after restoration of gauge invariance, we have reached an interesting 
agreement with production cross sections and polarisation
measurements by  CDF~\cite{CDF7997a,CDF7997b,Affolder:2000nn,Acosta:2001gv} 
and PHENIX~\cite{Adler:2003qs}.

In the case of $J/\psi$ and $\Upsilon(1S)$, we have used the leptonic decay
width to fix the normalisation of these vertex functions. This procedure 
has appeared as robust and reliable for 1S states~\cite{these} and is explained in the following. 
However in the case of radially excited states, 
an important feature has emerged  from calculations: the decay width, and thus 
the normalisation, are strongly dependent on the position of the node appearing in the vertex function.
This ambiguity in the determination of the decay width might also exist for hadronic decays, such as
$\psi'\to \rho \pi$, and is perhaps closely related to the $\rho-\pi$ puzzle.

\section{Our phenomenological approach}

Our approach to build the 
quarkonium vertex functions does not rely on the 
Bethe-Salpeter equation~\cite{Salpeter:1951sz} (BSE), 
usually used to constrain 
the properties of vertex functions; for example, to relate the mass of the bound 
state  to the mass of its constituent quarks. Indeed, besides problems with gauge invariance,
all predictions coming from BSE are realised within Euclidean space 
(see \eg~\cite{Maris:2003vk,Burden:1996nh,Ivanov:1998ms})
and we have shown that the continuation to Minkowski space can be problematic~\cite{these}.

We have therefore chosen to describe the quarkonia by a phenomenological 
vertex function and decided
to remain in Minkowski space, at variance
with other phenomenological models (see \eg~\cite{Ivanov:2000aj,Ivanov:2003ge}).
The price to pay for not using BSE is an additional uncertainty due to the functional
dependence of the vertex. Fortunately, it can be cancelled if one fixes
the normalisation from the study of the leptonic decay width, which is our main concern here.

\subsection{Choice for the vertex function}

It has been shown elsewhere (see \eg~\cite{Burden:1996nh}) that other Dirac structures 
than $\gamma^\mu$ for vector bound state are suppressed by one order of magnitude 
in the case of light mesons, and that this suppression is increased for the $\phi$ meson. 
Therefore their effect is expected to be even more negligible in heavy quarkonia.

As a consequence, the following phenomenological Ansatz for the vector meson vertex, 
inspired by spin-projection operators, is likely to be sufficient for our purposes. It
reads:
\begin{eqnarray}\label{vf}
  V_{\mu}(p,P) &=& \Gamma(p,P) \gamma_\mu,
\end{eqnarray}
with $P$ the total momentum of the bound state, $P=p_{1}-p_{2}$, and $p$ the
relative one, $p=(p_{1}+p_{2})/2$ as drawn on \cf{fig:BS_vertex_phenoa}. This Ansatz
amounts to multiplying the {\it point} vertex (corresponding to a structureless particle) 
by a function, $\Gamma(p,P)$.

\begin{figure}[!h]
\centering
\includegraphics[width=10.0cm]{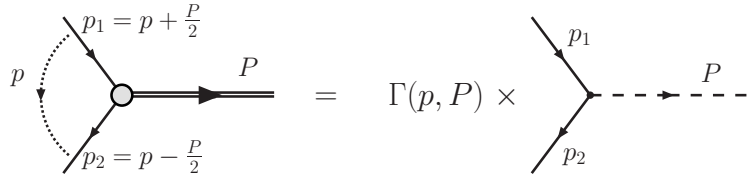}
\caption{Phenomenological vertex obtained by multiplying a {\it point vertex}, representing a structureless
particle, by a vertex function (or form factors).}
\label{fig:BS_vertex_phenoa}
\end{figure}

The function $\Gamma$, which is  called the 3-point vertex function, 
can be chosen in different ways. Two simple Ans\"atze are commonly used 
in the context of BSE. We consider both. 
They correspond to two extreme choices at large distances: a dipolar 
form which decreases gently with its argument, and a gaussian form: \index{vertex function!dipole}
\index{vertex function!gaussian}
\eqs{
\Gamma_0(p,P)=\frac{N}{(1-\frac{p^2}{\Lambda^2})^2} \text{ and } \Gamma_0(p,P)=N e^{\frac{p^2}{\Lambda^2}} 
,}
both with a free size parameter $\Lambda$. As $\Gamma(p,P)$ in principle
depends on $p$ and $P$, we may shift the variable and use $p^2-\frac{(p.P)^2}{M^2}$ instead
of $p^2$ which has the advantage of reducing to $-|\vect p|^2$ in the rest frame of the bound state.  
We refer to it as the {\it vertex function with shifted argument}. In the latter
cases, we have --in that frame--, 

\eqs{\label{eq:vertex_functions}
\Gamma(p,P)=\frac{N}{(1+\frac{|\vect p|^2}{\Lambda^2})^2} \text{ and }\Gamma(p,P)=N e^{\frac{-|\vect p|^2}{\Lambda^2}}.}

\subsection{Excited states}\index{excited states}

As is well-known, 
the number of nodes in the wave function, in whatever space, increases 
with the principal quantum number $n$. This simple feature can be used
to differentiate between $1S$ and $2S$ states.

We thus have simply to determine the position of the node of the wave function in momentum space.
To what concerns the vertex function, working in the meson rest frame, 
the node comes through a
prefactor, $1-\frac{|\vect p|}{a_{node}}$, which multiplies
the vertex function for the $1S$ state. Explicitly, $\Gamma_{2S}(p,P)$, for a node $a_{node}$, reads
\eqs{N'\left(1-\frac{|\vect p|}{a_{node}}\right)
\frac{1}{(1+\frac{|\vect p|^2}{\Lambda^2})^2} \text{ and  }
N'\left(1-\frac{|\vect p|}{a_{node}}\right) e^{\frac{-|\vect p|^2}{\Lambda^2}}.}

In order to determine the node position in momentum space, we can use two methods. The first is
to fix $a_{node}$  from its known value in position space, \eg~from potential studies, and to 
Fourier-transform the vertex function. In the case of a 
gaussian form, this can be carried out analytically. 

The second method is to impose the following relation\footnote{inspired by the orthogonality
between the $1S$ and $2S$ wave functions.} between the $1S$ and $2S$ vertex functions:
\eqs{\label{eq:ortho_1S_2S}
\int|\vect p|^2 d|\vect p| e^{\frac{-|\vect p|^2}{\Lambda^2}} \left(1-\frac{|\vect p|}{a_{node}}\right) e^{\frac{-|\vect p|^2}{\Lambda^2}}
=0.
}
The two methods give compatible results. 

\section{Normalising: the leptonic decay width}

The width in terms of the decay amplitude $\cal M$, is given by
\eqs{\label{eq:lept_width}
\Gamma_{\ell\ell}=\frac{1}{2 M}\frac{1}{(4\pi^2)}\int \left|\bar{\cal M}\right|^2 d_2(PS),
}
where $d_2(PS)$ is the two-particle phase space~\cite{barger}.
\index{two-particle phase space}

The amplitude is obtained as usual through Feynman rules, for which we 
use our vertex function at the meson-quark-antiquark vertex. At leading order,
the square of the amplitude is obtained from the cut-diagram drawn in \cf{fig:decay_diag_1}.

\begin{figure}[h]
\centering{\mbox{\includegraphics[width=8cm]{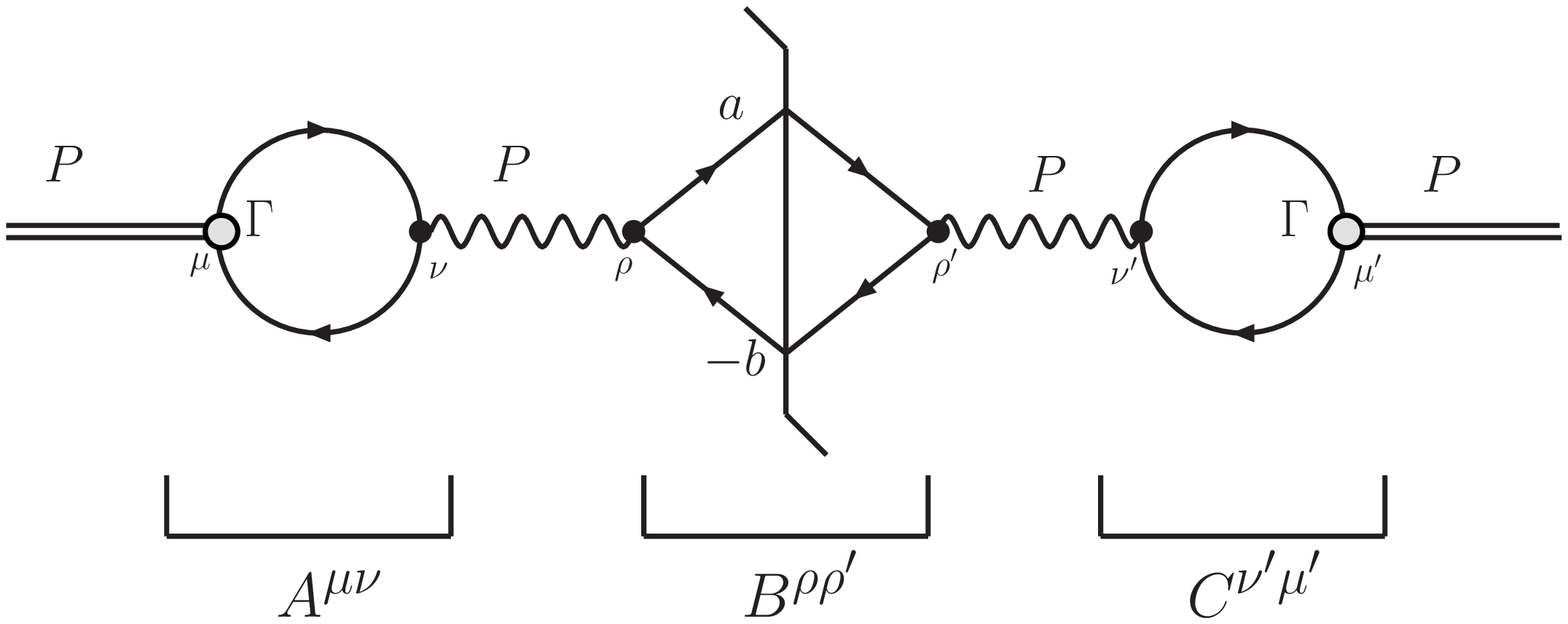}}}
\caption{Feynman diagram for $^3S_1\to \ell \bar \ell$.}\label{fig:decay_diag_1}
\end{figure}

\noindent In terms of the sub-amplitudes $A^{\mu\nu}$, $B^{\mu\nu}$ and $C^{\mu\nu}$ defined in
\cf{fig:decay_diag_1}, we have\footnote{In the Feynman gauge. The calculation can be shown
to be gauge-invariant, though.}:
\eqs{
\label{eq:decomp_ampl_inv_decay}
\int \left|\bar{\cal M}\right|^2 d_2(PS)=\frac{1}{3} 
\Delta_{\mu\mu'}
A^{\mu\nu}\left(\frac{-ig_{\nu\rho}}{M^2}\right)B^{\rho\rho'}(\frac{-ig_{\rho'\nu'}}{M^2})
C^{\nu'\mu'},
}
where the factor $\Delta_{\mu\nu}=(g_{\mu\nu}-\frac{P_\mu P_{\nu'}}{M^2})=
\sum_i \ep_{i,\mu} \ep^\star_{i,\mu'}$ results from the sum over polarisations 
of the meson and the factor $\frac{1}{3}$ accounts for the averaging on these initial polarisations.

\subsection{Sub-amplitude calculation}

To what concerns the sub-amplitude $B^{\rho\rho'}$, from the Feynman rules, and after 
integration on the two-particle phase space, we have
\eqs{
B^{\rho\rho'}=& (ie)^2 \left[\pi M^2 g^{\rho \rho'}-8 \frac{\pi M^2}{24}\left(g^{\rho\rho'}+2 
\frac{P^\rho P^{\rho'}}{M^2}\right) \right]=(ie)^2 \frac{2\pi}{3} M^2\underbrace{\left[g^{\rho\rho'}- 
\frac{P^\rho P^{\rho'}}{M^2}\right]}_{\Delta^{\rho\rho'}}.
}

For $A^{\mu\nu}$, from the Feynman rules and using the vertex functions discussed 
above, we have (see \cf{fig:decay_diag_1a})
\eqs{\label{eq:Amunu_start}
iA^{\mu\nu}&=-3e_Q \int\! \frac{d^4k}{(2 \pi)^4}
{\rm Tr} \left((i \Gamma(k,P)\gmu)  \frac{i(\ks k -\frac{1}{2}\ks P +m)}{(k-\frac{P}{2})^2-m^2+i\ep}
(i e \gnu)  \frac{i(\ks k +\frac{1}{2}\ks P +m)}{(k+\frac{P}{2})^2-m^2+i\ep}
\right)\\
 &=-3e_Q\int\! \frac{d^4k}{(2 \pi)^4}
 \Gamma(k,P)
\frac{g^{\mu\nu} (M^2+4m^2-4k^2)+8 k^\mu k^\nu-2P^\mu P^\nu}
{((k-\frac{P}{2})^2-m^2+i\ep)((k+\frac{P}{2})^2-m^2+i\ep)},
}
$e_Q$ is the heavy quark charge, $-1$ comes for the fermionic loop, $3$ is the colour factor.

\begin{figure}[h]
\centering{\mbox{\includegraphics[width=4cm]{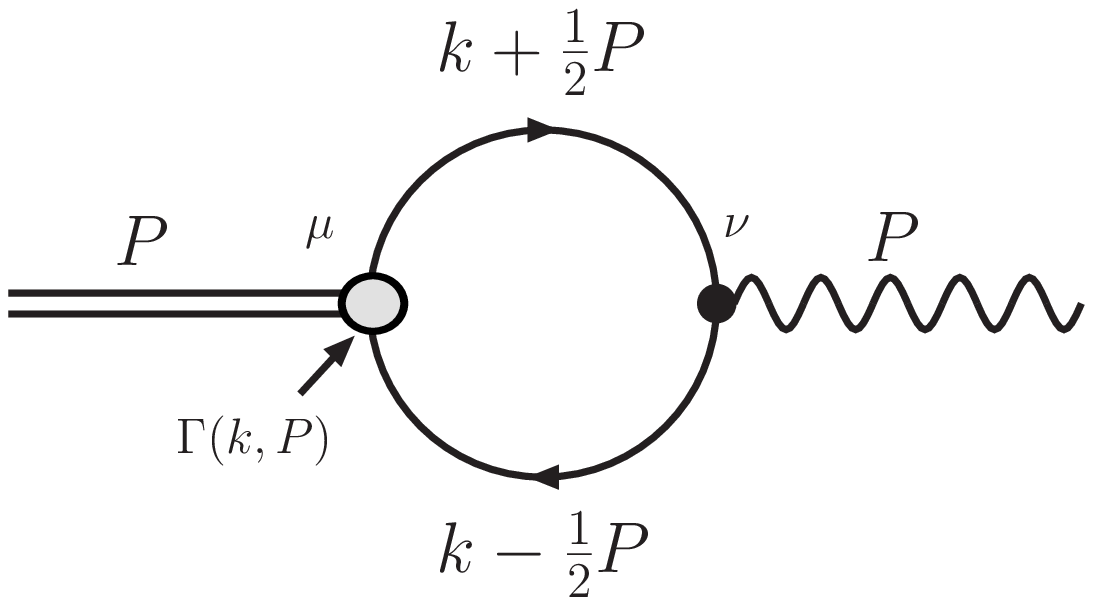}}}
\caption{Feynman diagram for $^3S_1\to \gamma^\star$.}\label{fig:decay_diag_1a}
\end{figure}

We can easily motivate our choice of the argument of the vertex function, 
beside its simple relation with that of wave functions. Its main virtue is to regularise
the integration in all spatial directions. In the meson rest frame, 
the $k^0$ integral can be done by standard residue techniques, and the remaining $\vect k$
integral is guaranteed to converge. If one uses $k^2$ instead of $|\vect k|^2$  as an argument, 
it is possible to show that along the light cone, one obtains logarithmic divergences,
which would presumably need to be renormalised. Besides, the $k^0$ integral then also 
becomes dependent on the singularity structure of the vertex function. These two reasons
make us prefer the shifted vertex in our calculation.

A notable simplification 
can be obtained by guessing the tensorial form of $A^{\mu\nu}$. Indeed, current
conservation (gauge invariance) for the photon can be expressed as:
\index{integration by residues}
\index{current conservation}\index{gauge!invariance}

\eqs{\label{eq:GI_Amunu}
A^{\mu\nu}P_\nu=0.
}
It is equivalent to 
\eqs{
iA^{\mu\nu}=iF (g^{\mu\nu}-\frac{P^\mu P^\nu}{M^2})=iF\Delta^{\mu\nu},
}
the coefficient $F$ being $\frac{A^\mu_{\ \mu}}{3}$ since $A^\mu_{\ \mu}= A^{\mu\nu}g_{\nu\mu}=3F$.
This assumption can be easily verified in the bound-state rest frame $P=(M,0,0,0)$ 
for which \ce{eq:GI_Amunu} reduces
to $A^{\mu0}M=0\Rightarrow A^{\mu0}=0$. We shall check this at the end of this calculation.

It is therefore sufficient to compute the following quantity, where we set $|\vect k|^2\equiv K^2$ and
define $\Gamma(-K^2)=\Gamma(k,P)$,
\eqs{
iA^\mu_{\ \mu}=-3e_Q&\int_0^\infty 4 \pi K^2 dK \Gamma(-K^2)\times \\
&\int_{-\infty}^\infty \frac{dk_0}{(2 \pi)^4}
\frac{4 (-2(k^2-\frac{M^2}{4})+4m^2)}
{((k-\frac{P}{2})^2-m^2+i\ep)((k+\frac{P}{2})^2-m^2+i\ep))}.
}

Let us first integrate on $k_0$ by residues. Defining $E=\sqrt{K^2+m^2}$, we determine
the position of the pole on $k_0$ (still in the bound-state rest frame) as
\eqs{
&(k\pm\frac{P}{2})^2-m^2+i\ep=(k_0\pm\frac{M}{2})^2-E^2+i\ep.
}

To calculate the integral on $k_0$, we choose the contour as drawn in \cf{fig:k0_plane}.
Two poles $-E-\frac{M}{2}$ and $-E+\frac{M}{2}$ are located in the upper half-plane and 
the two others in the lower
half-plane. We shall  therefore have two residues to consider. The contribution of the 
contour ${\cal C}_R$ vanishes as $R$ tends to $\infty$. 

\eqs{
&iA^\mu_{\ \mu}=\frac{-3e_Q}{\pi^3} \int_0^\infty K^2 dK \Gamma(-K^2)\times\\
&2i\pi\left[
\frac{-2(-E-\frac{M}{2})^2+\frac{M^2}{2}+2E^2+2m^2}{(-2E)(-M)(-2(E+\frac{M}{2}))}\right.
+\left.
\frac{-2(-E+\frac{M}{2})^2+\frac{M^2}{2}+2E^2+2m^2}{M(-2(E-\frac{M}{2}))(-2E)}
\right]\\
&=\frac{-3e_Q}{\pi^3} \int_0^\infty K^2 dK \Gamma(-K^2)\frac{2i\pi}{4ME}
\underbrace{\left[-\frac{2EM+2m^2}{E+\frac{M}{2}}+\frac{2EM+2m^2}{E-\frac{M}{2}}\right]}
_{\frac{M(4E^2+2m^2)}{E^2-\frac{M^2}{4}}}.
}

\begin{figure}[ht]
\centering{\mbox{\includegraphics[width=12 cm]{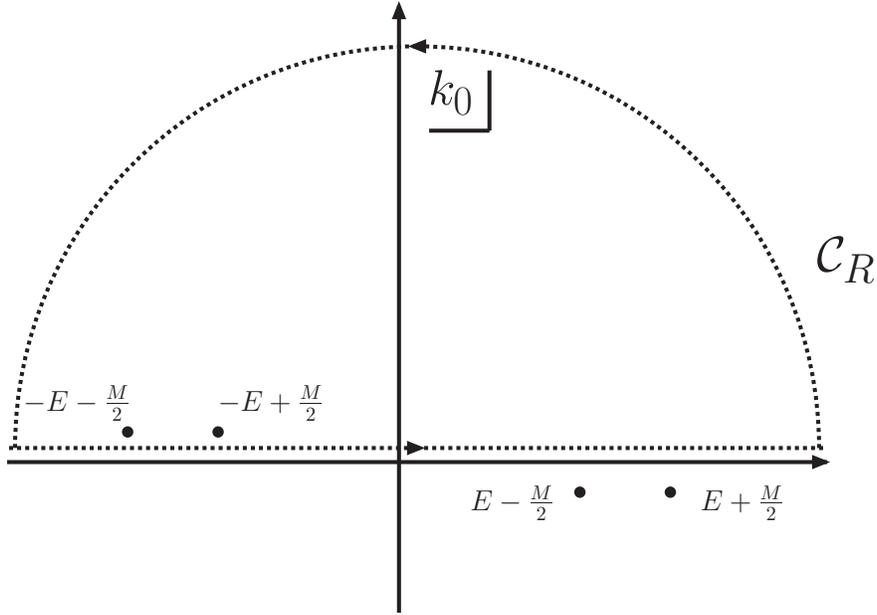}}}
\caption{Illustration of the contour chosen to integrate on $k_0$.}\label{fig:k0_plane}
\end{figure}

The procedure is correct only if the poles of the upper-plane are on the left
side and those of the lower-plane are on the right\footnote{If they cross, the integral
acquires a discontinuity from the pinch and becomes complex.}.

This crossing would first occur between the points $-E+\frac{M}{2}$ and $E-\frac{M}{2}$. 
Solving for $E$, we have:
\eqs{
-E+\frac{M}{2}=E-\frac{M}{2}  &\Rightarrow    \frac{M^2}{4}-m^2=K^2>0.
}
The crossing is therefore impossible when $M<2m$. Therefore, to get
a coherent description of all charmonia (resp. bottomonia) below the open charm (resp. beauty) 
threshold, we shall set the quark mass high enough to avoid crossing for all of these.  This sets
$m_c$ to 1.87 GeV ($m_D$) and $m_b$ to 5.28 GeV  ($m_B$). Considering the variety of results
obtained from potential models, this seems to be a sensible choice.\index{B@$B$}\index{D@$D$}

Putting	it all together, we get 
\eqs{
A^\mu_{\ \mu}=\frac{-3 e_Q}{\pi^2} \int_0^\infty dK \frac{K^2\Gamma(-K^2)}{\sqrt{K^2+m^2}}
\frac{(2K^2+3m^2)}{(K^2+m^2-\frac{M^2}{4})}.
}
One is left with the integration on $K$ for which we define the integral $I$ depending 
on the vertex function whose normalisation is pulled off,
\eqs{\label{eq:I}
I(\Lambda,M,m)\equiv\int_0^\infty  \frac{dK K^2}{\sqrt{K^2+m^2}}
\frac{\Gamma(-K^2)}{N}
\frac{(2K^2+3m^2)}{(K^2+m^2-\frac{M^2}{4})},
}
$I$ is a function of $\Lambda$ through the vertex function $\Gamma(-K^2)$ 
and is not in general computable analytically.
In the following we shall leave it as is and express 
$A^{\mu\nu}$ as:
\eqs{
A^\mu_{\ \mu}=\frac{-3 e_Q}{\pi^2} N I(\Lambda,M,m) \Rightarrow A^{\mu\nu}=\frac{- e_Q}{\pi^2} N I(\Lambda,M,m) \Delta^{\mu\nu}.
}

Finally, to what concerns the sub-amplitude $C$, we simply have 
\eqs{
C^{\nu'\mu'}=(A^{\nu'\mu'})^\dagger
=A^{\nu'\mu'}.
}

\subsection{Results}

Now that all quantities in \ce{eq:decomp_ampl_inv_decay} are determined, we can combine them. 
Hence, we obtain\footnote{Recall that the projector $\Delta_{\mu\nu}$ satisfies 
$\Delta_{\mu\nu}\Delta^{\mu\nu}=3$ and $\Delta_{\mu\nu}\Delta^{\mu\nu'}=\Delta_{\nu}^{\ \nu'}$.}:
\eqs{&\int \left|\bar{\cal M}\right|^2 d_2(PS)=\frac{\Delta_{\mu\mu'}}{3}
\frac{-1}{M^4}\left(\frac{ e_Q}{\pi^2} N I(\Lambda,M,m)\right)^2\left((ie)^2 \frac{2\pi}{3} M^2\right)
  \Delta^{\mu\nu}g_{\nu\rho}
\Delta^{\rho\rho'}g_{\rho'\nu'}
\Delta^{\nu'\mu'}
}

The leptonic decay width eventually reads from~\ce{eq:lept_width}:
\eqs{\Gamma_{\ell\ell}=N^2 \frac{e^2}{12 \pi M^3}  \left(\frac{e_Q}{\pi^2} I(\Lambda,M,m)\right)^2.}

\subsubsection{Numerical results}

The value of $N$ is in practice obtained by replacing $\Gamma_{\ell\ell}$ by its measured value, $e_Q$ by
$\frac{2e}{3}$ for $c$ quark and $\frac{-e}{3}$ for $b$ quark
and, finally, by introducing the value obtained for $I$ for the 
chosen value of $\Lambda$. We sketch in \cf{fig:N_jpsi_lam} and \cf{fig:N_dip_upsi_lam} 
some plots of $N$ for the $J/\psi$ and for the $\Upsilon(1S)$ to show its dependence 
on $\Lambda$ and $m_Q$ and the value it actually takes. 

\begin{figure}[h!]
\centering{\mbox{\includegraphics[width=0.49\textwidth]{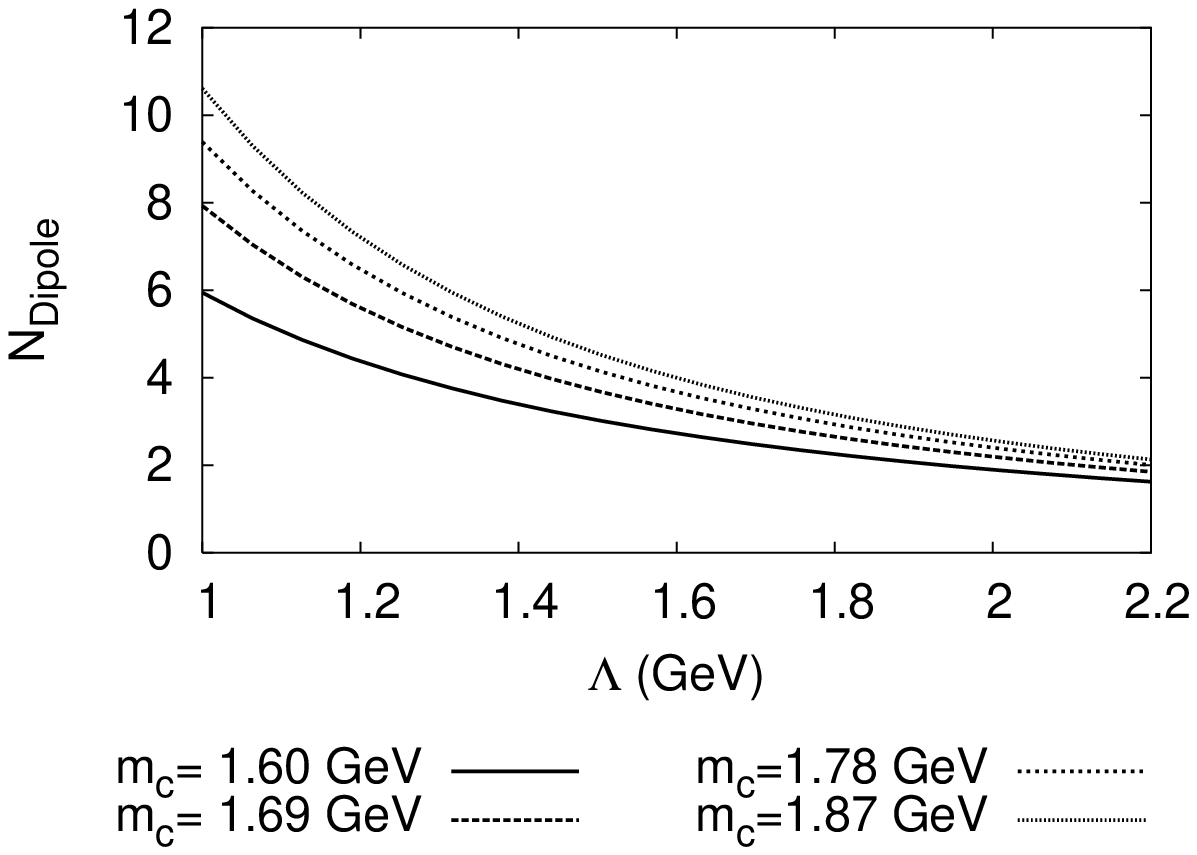}
\includegraphics[width=0.49\textwidth]{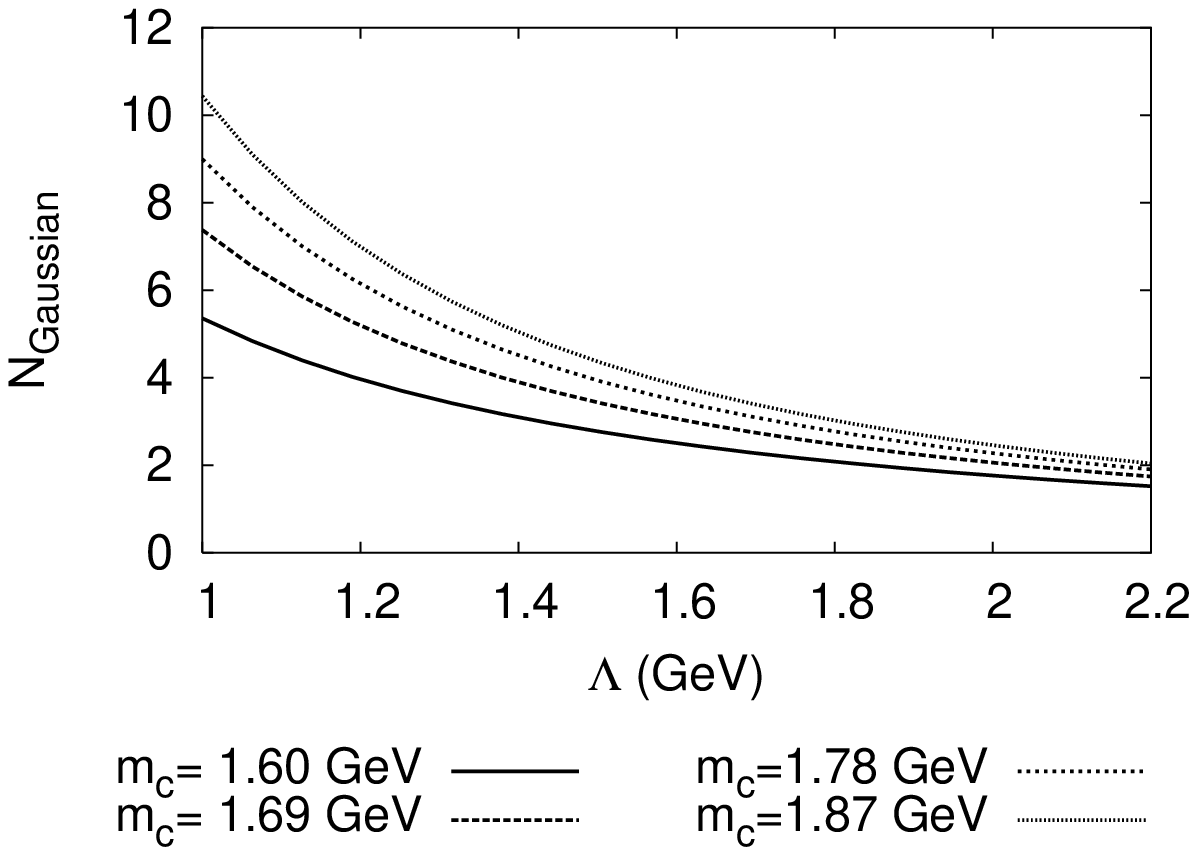}}}
\caption{Normalisation for a dipolar (resp. gaussian) form for $J/\psi$ as a function of $\Lambda$: right (resp. left ).}
\label{fig:N_jpsi_lam}
\end{figure}

\begin{figure}[h!]
\centering{\mbox{\includegraphics[width=0.49\textwidth]{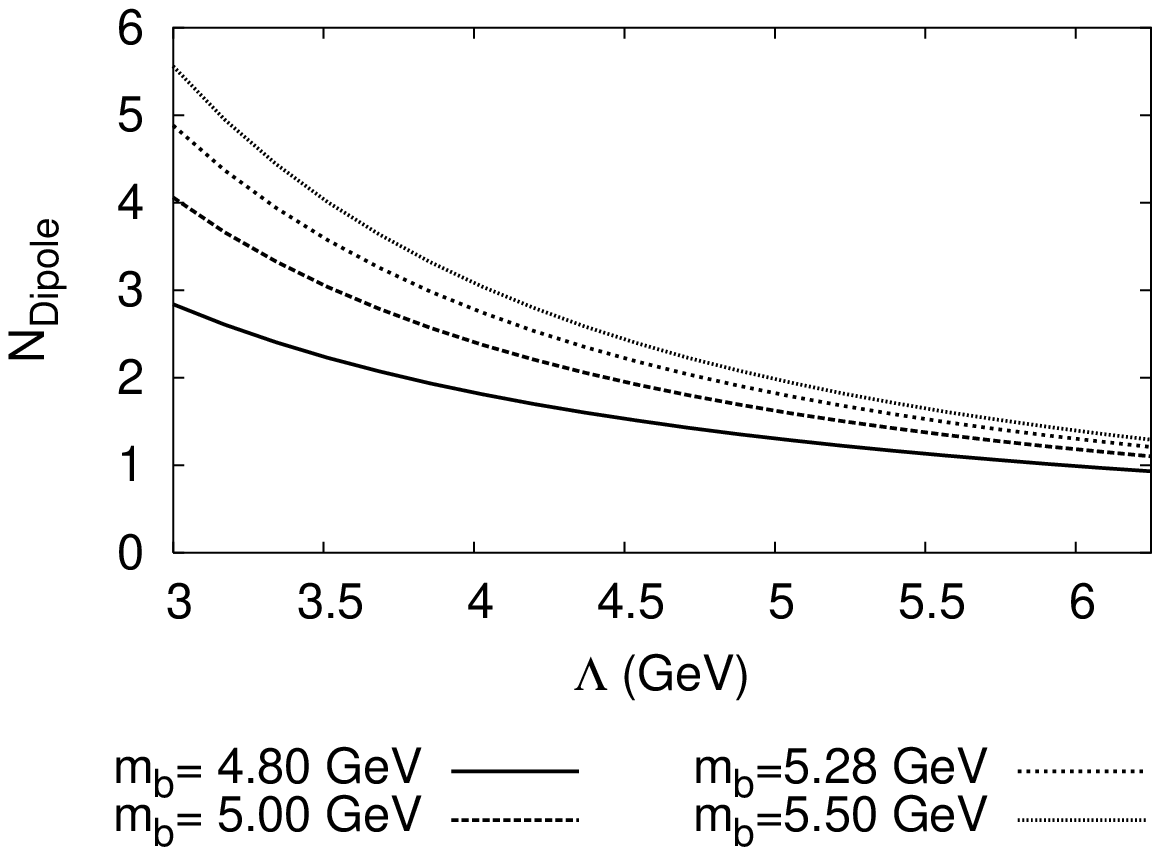}
\includegraphics[width=0.49\textwidth]{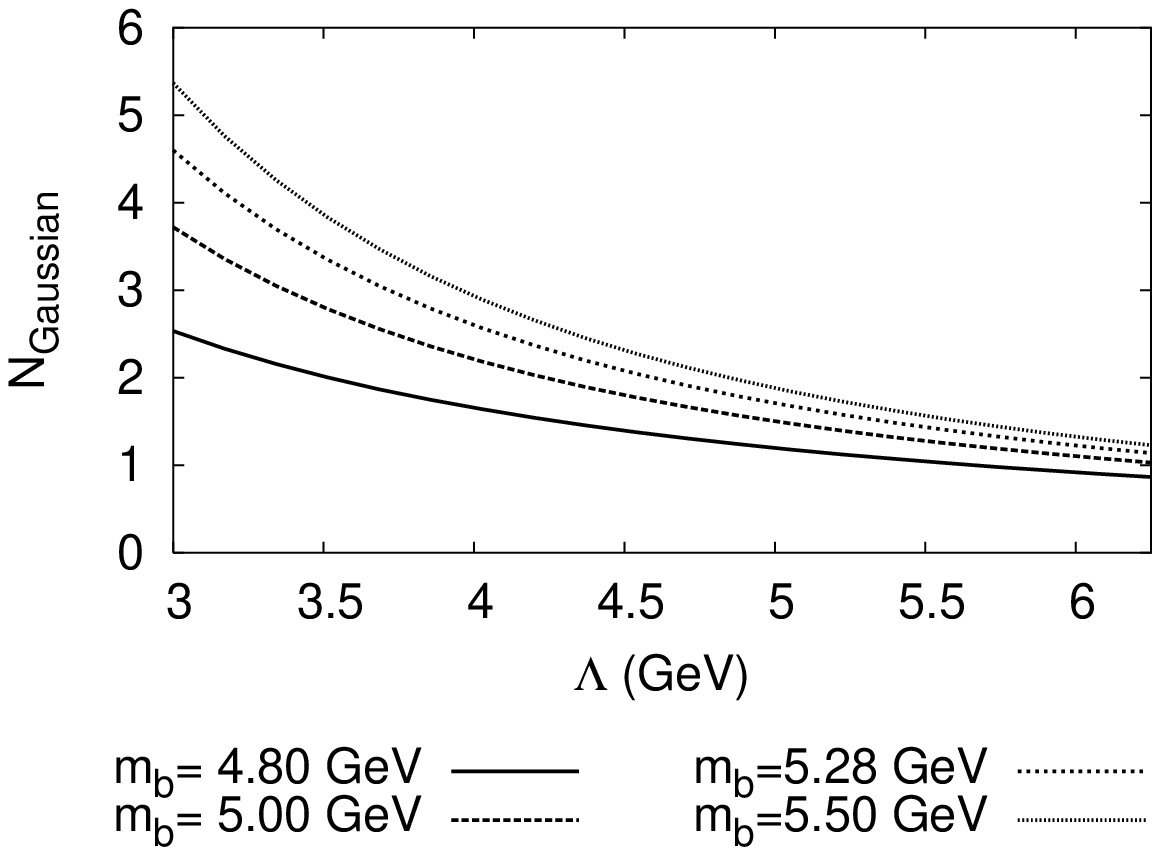}}}
\caption{Normalisation for a dipolar (resp. gaussian) form for $\Upsilon(1S)$  as a function of $\Lambda$: right (resp. left ).}
\label{fig:N_dip_upsi_lam}
\end{figure}

\begin{table}[h!]
\begin{tabular}{c||c}
\hline $m_c$ (GeV) & Vertex functions \\ 
\hline 
\begin{tabular}{c}
     \\
     \\
1.6  \\
1.7  \\
1.87 \\
\end{tabular}
&
\begin{tabular}{c|c}
Gaussian& Dipole \\
\hline
\begin{tabular}{c|c} 
$N$     & $\Lambda$ (GeV)  \\
 1.59   &  2.34       \\
  2.28   &  2.08    \\
 3.31   &  1.97       \\
\end{tabular}
&
\begin{tabular}{c|c} 
$N$     & $\Lambda$ (GeV)  \\
 1.33   &   2.87      \\
 1.87   &   2.54     \\
 2.76   &   2.31       \\
\end{tabular}         \\
\end{tabular} \\
\hline
\end{tabular}
\caption{Set of values for $m_c$, $\Lambda$ and $N$ obtained for the $J/\psi$ within BSE approach~\cite{costa}.}
\label{tab:res_norm_meth2}
\end{table}

We have chosen the range of size of the mesons, or $\Lambda$, following BSE studies~\cite{costa} and other 
phenomenological models~\cite{Ivanov:2000aj,Ivanov:2003ge}, where
 the commonly accepted values for the $J/\psi$ are from around $1$ to $2.4$ 
GeV, and for the $\Upsilon(1S)$ from around $3.0$ to $6.5$ GeV. For illustration, we put in~\ct{tab:res_norm_meth2} 
several values of $\Lambda$ obtained with BSE~\cite{costa}, accompanied with the vertex-function normalisation.

\subsubsection{Result for $\psi'$ and implication for the $\rho$-$\pi$ puzzle}

To what concerns the $\psi'$, we give the results for $\Lambda$=1.8 GeV, $m_c=1.87$ GeV 
 and a gaussian vertex function (see~\cf{fig:normalisation_psip_290305}). The normalisation $N'$
diverges for $a_{node} \simeq 1.35$ GeV. This comes from the cancellation of the leptonic
decay width for this value of $a_{node}$.

\begin{figure}[h!]
\centering{\mbox{\includegraphics[width= 8cm]{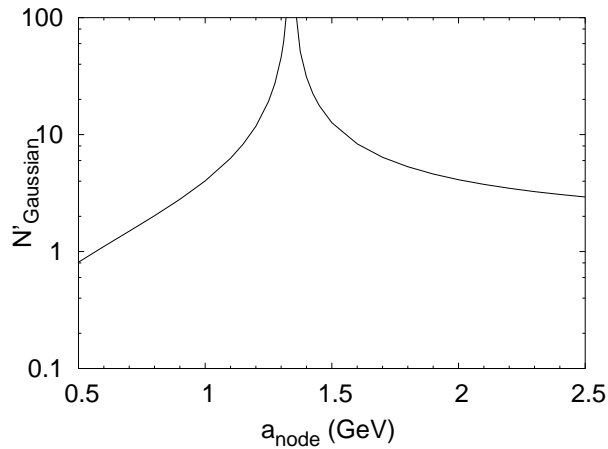}}}
\caption{$N'$ for the $\psi'$ as a function of $a_{node}$.}
\label{fig:normalisation_psip_290305}
\end{figure}

This cancellation can be traced back to the integrand
of $I$ (see \ce{eq:I}), which we note  $\frac{dI}{dK}$ (see~\cf{fig:dIdk}). 
The cancellation of the positive contribution by the negative one therefore occurs in this case for  $a_{node}=1.35 $ GeV. 

\begin{figure}[h!]
\centering{\mbox{\includegraphics[width=8 cm]{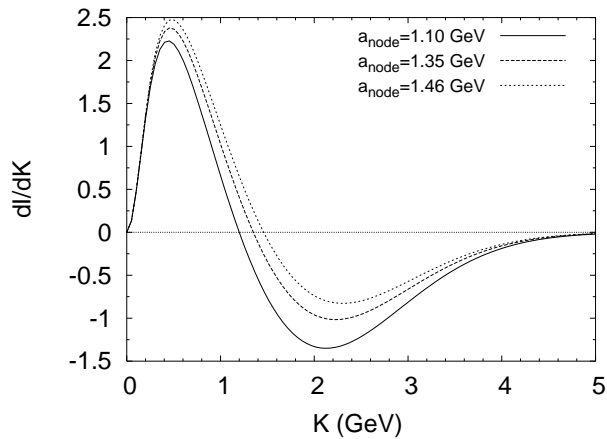}}}
\caption{Evolution of the integrand in $I$ as a function of $K$ and for three values of $a_{node}$ in the
$\psi'$ case.}\label{fig:dIdk}
\end{figure}

The constraint for the node position~\ce{eq:ortho_1S_2S} gives $a_{node}=1.46 $ GeV for
these values of $\Lambda$ and $m_c$. For this value of $a_{node}$, the normalisation $N'$ is 16.36.  
This was the value that we have retained for the numerical applications of~\cite{these}. 

In the case of the hadronic decay $\psi'\to \rho \pi$, we expect it to proceed via an off-shell
$\omega$ (see \cf{fig:psip_rho_pi}). The integral $I$ involving the $\psi'$ vertex function
will be slightly different and might be, for the same value of $a_{node}$, drastically suppressed.

\begin{figure}[ht]
\centering{\mbox{\includegraphics[width=6 cm]{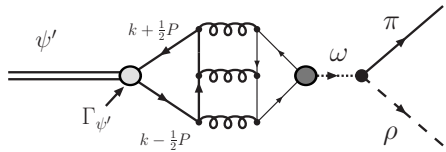}}}
\caption{Possible decay of a $\psi'$ into a $\rho$ - $\pi$ pair via an off-shell $\omega$.}
\label{fig:psip_rho_pi}
\end{figure}

We can therefore expect a severe suppression of the decay amplitude 
compared to the nodeless, \ie~$J/\psi$, case. This suppression could in turn explains 
the $\rho-\pi$ puzzle, namely that the measured 
ratio $\frac{\Gamma(\psi'\to \rho\pi)}{\Gamma(J/\psi\to \rho\pi)}$ 
is not of the order of 15 \% (expected from the leptonic decays) 
but rather smaller than 1\%.

\section{Conclusion}

We have explained our approach to describe on a phenomenological basis the internal dynamics of
heavy quarkonia in the context of a Feynman-diagram calculation. We have also provided a way to
represent the distinct features of $2S$ states, like the $\psi'$. We have seen that 
a simple but robust constraint on the vertex functions  for $1S$-states could be achieved by imposing that 
their normalisation reproduces the leptonic decay width, through a simple leading order calculation. 

An interesting simplification can be also obtained by shifting the argument of the vertex function, 
namely the relative momentum of the quark inside the quarkonium, into a quantity that reduce to 
the tri-dimensional relative momentum in the meson rest frame. This enable us to work out 
analytically, for whatever vertex function, the integration on $k_0$.

Furthermore, we have shown that the leptonic decay width was very dependent on
the node position, which, incidentally, is not a completely constrained parameter. This 
induces the same effects on the normalisation (see~\cf{fig:normalisation_psip_290305}). 
This has consequences on production processes where our approach to describe internal dynamics of
heavy quarkonia can be applied (see~\cite{these,Lansberg:2005pc,article2}). It should be interesting to see 
whether this happens for other excited states. Finally, we suggest that this feature 
typical of radially excited states could be the awaited explanation for the 
longtstanding $\rho-\pi$ puzzle.


\begin{theacknowledgments}
This work is supported by a IISN (Interuniversity Institute of Nuclear Science, Belgium) Research 
Fellowship and has been done in collaboration with J.R.~Cudell and Yu.L. Kalinovsky. 
We would also like to gratefully  thank Yu.L. Kalinovsky and P. Costa for sharing their 
results from the BSE approach.
\end{theacknowledgments}



\bibliographystyle{aipprocl} 





\end{document}